# Cation Dominated but Negatively Charged Na$_2$SO$_{4,aq}$-Graphene Interfaces

Ademola Soyemi and Tibor Szilvási*

Department of Chemical and Biological Engineering, The University of Alabama, Tuscaloosa, Alabama 35487, United States

* Email: tibor.szilvasi@ua.edu

## Abstract

The distribution of ions and their impact on the structure of electrolyte interfaces plays an important role in many applications. Interestingly, recent experimental studies have suggested the preferential accumulation of SO$_4^{2-}$ ions at the Na$_2$SO$_{4,aq}$-graphene interface in disagreement with the generally known tendency of cations to accumulate at graphene-electrolyte interfaces. Herein, we resolve the atomistic structure of the Na$_2$SO$_{4,aq}$-graphene interfaces in the 0.1-2.0 M concentration range using machine learning interatomic potential-based simulations and simulated sum frequency generation (SFG) spectra to reveal the molecular origins of the conundrum. Our results show that Na$^+$ ions accumulate between the outermost and second water layers whereas SO$_4^{2-}$ ions accumulate within the second interfacial water layer indicating cation dominated interfaces. We find that the interfacial region (within ~10 Å of the graphene sheet) is negatively charged due to sub-stoichiometric Na$^+$/SO$_4^{2-}$ ratio at the interface. Our simulated SFG spectra show enhancement and a red-shift of the spectra in the hydrogen bonded region as a function of Na$_2$SO$_4$ concentration similar to measurements due to SO$_4^{2-}$-induced changes in the orientational order of water molecules in the second interfacial layer. Our study demonstrates that ion stratification and ion-induced water reorganization are key elements of understanding the electrolyte-graphene interface.

## Introduction

The aqueous electrolyte-graphene interface plays an important role in numerous applications such as electrochemistry/electrocatalysis,[1, 2] storage devices,[3] and water treatment[4] among others. The behavior of the electrolyte at the interface greatly influences phenomena such as the formation of the electrical double layer (EDL), charge transfer, and chemical reactions which determine the performance and stability of electrochemical devices. Thus, understanding the fundamental nature of the interface and how dissolved ions affect the molecular structure of the interface is essential to improving the performance of electrochemical devices and facilitating scientific advancements.

Previous experimental and computational studies have explored various electrolyte-graphene interfaces.[5-11] The surface-specific sum frequency generation (SFG) spectroscopy has commonly been used to characterize the structure of the graphene-electrolyte interface,[10, 12-14] however, interpreting SFG spectra can be difficult due to low signal-to-noise ratio of experiments,[15] substrate effects,[9, 16] and lack of phase-resolution in the SFG spectra typical of homodyne-detected SFG measurements.[17] To overcome the challenges in understanding the molecular origin of the SFG spectra, machine learning interatomic potentials (MLIPs) trained on density functional theory (DFT) data have recently emerged as a potential solution for interpreting SFG spectra,[18-20] circumventing the accuracy issues of classical force fields[8, 21, 22] and the time- and length-scale limitations of DFT.[23] It has been shown that the stratification of ions into different interfacial water layers, driven by differences in the propensity of cation and anions to accumulate at the graphene-water interface, leads to ion-induced orientational reorganization of water that manifests as spectral changes in the SFG spectra.[5, 8, 24] However, the aforementioned studies have mainly focused on simple monovalent ion pairs such as NaCl,[6, 24] H$_3$O$^+$/OH$^-$,[5] or NaClO$_4$[9] while more complex solutions containing divalent-monovalent ion pairs (or 2:1 electrolyte systems) are poorly understood. To address this, Yang et al.[10] recently measured the SFG spectra for the Na$_2$SO$_{4,aq}$- graphene interface and showed an enhancement of SFG spectra in the hydrogen bonded region (i.e., between 3000 and 3400 cm$^{-1}$) with increasing concentration.[10] Here, it was suggested that the spectral changes were due to accumulation of

$SO_4^{2-}$ ions near the interface, though previous studies have suggested different ion partitioning scenarios at graphene-electrolyte interfaces.[7, 24, 25]

In this computational study, we provide a molecular interpretation of the observed spectral changes at the $Na_2SO_{4,aq}$-graphene interface using MLIP-based MD simulations and reveal how the apparent conundrum in the interpretation of previous measurements can be resolved. Our results show that the $Na^+$ and $SO_4^{2-}$ ions are stratified whereby the $Na^+$ ions accumulate between the outermost (first) and second interfacial water layers (~5 Å from the graphene) while $SO_4^{2-}$ ions accumulate within the second interfacial water layer (~7 Å from the graphene sheet). We show across a bulk concentration of 0.1 to 2.0 M that $Na^+$ ions dominate the interface in terms of surface concentration (0.3 to 5.3 M) compared to $SO_4^{2-}$ ions (0.2 to 4.8 M). However, the sub-stoichiometric $Na^+/SO_4^{2-}$ ratio at the interface provides a net negative charge at the interface in agreement with experiments[10]. Additionally, we show using layer-by-layer analysis of the SFG and water orientation that the SFG peak in the hydrogen bonding region intensifies as a function of $Na_2SO_4$ concentration and becomes red-shifted due to increasing interaction of water molecules with $SO_4^{2-}$ ions which induce orientational order in interfacial water molecules. Overall, our results emphasize that Na cations accumulate near graphene-water interfaces over sulfate anions while demonstrating that the ion stratification and ion-induced water reorganization at the interface is a critical element in understanding the electrolyte-graphene interface and interpreting SFG spectra.

**Methods**

**MLIP Training Methodology**

We performed all reference DFT calculations using the Vienna Ab initio Simulation Package (VASP) version 6.3.2.[26, 27] using the RPBE functional.[28] We represented the electron-core interaction using projector augmented waves (PAWs) with a plane wave basis energy cutoff of 520 eV. We performed single-point calculations using gamma-centered k-point sampling of the Brillouin zone and broadened the orbital energies using a Gaussian smearing width of 0.03 eV. We set the energy convergence criterion to $10^{-7}$ eV to obtain accurate forces. Following previous studies[29-31] that have shown that dispersion-corrected RPBE accurately describes bulk water, we include dispersion corrections using Grimme's D3 method with zero damping.[32, 33] In addition, a recent study has shown that the D3 method inaccurately calculates the $C_{AB}^6$ term for pairs involving cations such as $Na^+$.[34] Consequently, we modified the DFT-D3 code to exclude the contributions of $Na^+$ and recalculated the D3 contribution in all structures before MLIP training.

We train MLIPs using NequIP package[35] that we find to be easier to train than the Allegro package[36] for the graphene-electrolyte interface. The development branch of NequIP, downloaded on October 23, 2023 from GitHub (https://github.com/mir-group/nequip), is used for all training and inference. The initially generated dataset comprised 2400 structures. The initial dataset was split into training set (75%) and validation set (25%), and multiple NequIP [35] MLIPs with a 6 Å radial cutoff and equivariant E(3) products up to $L_{max}$ = 2 in the tensor layers were trained to find a MLIP with the optimal set of hyperparameters. We note that training set structures are large enough that the radial cutoff of our MLIP (6 Å) does not pass through the periodic boundaries. Consequently, we expect the MLIP to not become overfitted on the training structures and to transfer well to systems of larger sizes.[37] We trained the NequIP MLIPs in two stages of 500 epochs each using an initial learning rate of 0.02; first where the force to energy ratio in the loss function is 20:10, ensuring the forces are trained accurately, and second where the force to energy ratio is 10:10. In both stages, we set the stress loss coefficient very high (100,000) to ensure the stresses are also sufficiently learned. Seven of the best MLIPs, based on the validation loss, were then selected to be used as the committee of MLIPs (See our Zenodo repository, DOI: https://doi.org/10.5281/zenodo.15512423) for training input files for all committee members).

For training of the MLIP, we followed an iterative training approach whereby we systematically improved the MLIP over nine training iterations by incorporating new structures via a query-by-committee data selection method (see Figure S13). Recent work has shown that by following a model distillation approach,[38, 39] small, fast, and accurate potentials can be trained on data generated by a foundational potential. Therefore, following our previous work,[31] we generated initial data including all relevant subsystems such as bulk water, sodium sulfate in bulk water (at different concentrations), and slab systems such as water on graphene, sodium sulfate in water on graphene (at different concentrations) using the MACE-MP-0 (medium) pretrained MLIP[40] that has been shown to adequately model solid-liquid interfaces[41] via the Atomic Simulation Environment (ASE, version 3.22.1).[42] Our converged MLIPs required 9 iterations with a final training set of 4,047 structures and a validation set of 2,213 structures. Similarly for generating an independent test set, we run molecular dynamics simulations in the OpenMM[43] code (version 8.1.1) for all the subsystems using our final MLIP. The final MLIP achieved a test set energy, force, and stress MAE of 0.53 meV/atom, 29 meV/Å, 0.06 meV/Å$^2$, respectively. See Sections 1 to 5 of the Supporting Information for further details on data generation and detailed analysis of test set errors.

**MLIP-based Molecular Dynamics Simulations**

To study the effects of $Na_2SO_4$ on the graphene-water interface, we use the aqueous $Na_2SO_4$-graphene interface as a model and explore a range of $Na_2SO_4$ concentrations. As we showed recently,[44, 45] a bulk concentration of 1.0 M can be achieved in a $Na_2SO_4$-graphene system within two iterative quasi-constant chemical potential (iqCμMD) MD iterations.[44, 45] Therefore, we initialized the $Na_2SO_4$-graphene interface over a range of 0.1 to 2.0 M concentration, based on the total number of $Na_2SO_4$ molecules, in addition to the neat $Na_2SO_4$-graphene interface (0.0 M). As we show in Table S1, our iqCμMD-based approach allows us to study the $Na_2SO_4$-graphene interface across a range of bulk concentrations while maintaining constant concentration in the bulk region of the simulation cell. The simulation cell had dimensions of 34.15 Å x 29.57 Å x 110.00 Å comprising of 2 fixed graphene sheets with 384 carbons each (768 carbon atoms in total), 1,600 water molecules (4800 atoms in total, and ~49 Å in thickness), and 3 to 58 $Na_2SO_4$ molecules (Table S1) corresponding to a nominal concentration of 0.1 to 2.0 M (See Table S1). It should be noted that the graphene sheets were held fixed at 28.25 Å and 81.75 Å in the z direction and periodic boundary conditions were applied in the x, y, and z directions. All production MLIP-based MD simulations were performed using the OpenMM-ML plugin (version 1.1) in the OpenMM[43] code (version 8.1.1). For production runs, we created multiple replicates of the initial solution at each concentration using Packmol.[46] We optimized the system and then equilibrated within the NVT ensemble under Langevin dynamics using a 1 fs timestep at 310 K for 300 ps with the hydrogen mass set to 4 amu. After the initial equilibration, we reset the hydrogen mass to 1 amu and ran NVT simulations with a 1 fs timestep at 300 K for at least 4 ns to ensure efficient sampling of the bulk and interfacial regions. We emphasize that we simulated the $Na_2SO_4$-graphene system at each concentration for at least 50 ns cumulatively.

**SFG Methodology**

We calculated the resonant term of the second order non-linear susceptibility using the surface-specific velocity-velocity autocorrelation formalism as developed by Nagata and co-workers.[47, 48] Since our goal in this work is to capture the overall trend in the SFG spectra, we ignored intermolecular couplings terms which have been shown not to affect the SFG spectra significantly from a qualitative perspective.[47, 48] Thus, we calculated the averaged SFG spectra from five 500 ps NVT simulations using a 0.5 fs timestep. We emphasize that we initialized the simulation for each concentration from a converged MD trajectory. To generate the spectra, we computed the SFG spectra using the SFG-spectra-tool[47, 49] and applied the quantum correction factor and included non-Condon effects using the equation below:

$$\frac{Q(\omega)\mu'(\omega)\alpha'(\omega)}{i\omega^2} \quad (1)$$

where ω is the frequency, and the quantum correction factor ($Q(\omega)$) is defined as follows,

$$\frac{\beta\hbar\omega}{1-\exp(-\beta\hbar\omega)} \quad (2)$$

and the transition dipole moment $\mu'(\omega)$ and polarizability $\alpha'(\omega)$ were parametrized by Skinner and co-workers.[50, 51]

For the calculation of the SFG spectra, we took the interfacial region (~9 Å) to be the three interfacial regions (L1, L2, L3) as defined in Section 8 of the Supporting Information whereby the spectra was averaged across both graphene-water interfaces. Additionally, we computed the SFG contribution from the bulk region, while we approximate the third-order susceptibility by calculating the layer by layer contributions to the SFG spectra as suggested in previous studies.[19] To facilitate comparison with experimental results, our calculated spectra frequencies are scaled by a factor of 0.96 to approximate the impact of nuclear quantum effects in the high-frequency region, as suggested in previous studies.[19] We also note that various methods exist in literature[19, 52] for simulating SFG spectra and a comprehensive comparison of SFG simulation methods will be the subject of a future study.

**Results & Discussion**

To begin our analysis, we first address the accuracy of our local MLIP without explicit treatment of long-range Coulomb interactions. Some recent studies[53, 54] have suggested that explicit treatment of long-range electrostatics is required to recover the correct bulk properties of water in interfacial systems with and without ions. However, other studies have also shown that short-ranged MLIPs are accurate for modeling the graphene-water interface with or without ions.[5, 6, 12, 55, 56] Thus, we first demonstrate that our trained MLIP can recover correct interfacial and bulk structure. We then discuss the effect of $Na_2SO_4$ on the structure of water at the interface. We analyze the distribution of $Na^+$ and $SO_4^{2-}$ ions as a function of the distance from the graphene surface. Next, we study the SFG spectra as a function of the nominal bulk $Na_2SO_4$ concentration. To establish a direct link between the organization of water molecules in distinct layers and the SFG spectra, we also explore the SFG response in the water layers. To further elucidate the effect of the orientation of water molecules on the SFG spectra, we analyze water orientation in the different layers as a function of nominal $Na_2SO_4$ concentration. Lastly, we analyze the hydrogen bonding, tetrahedrality, and in-plane diffusivity in the water layers as a function of nominal $Na_2SO_4$ concentration.

In Figure 1, we analyze the average water dipole ($\theta_{DW}$, see Section 9 of Supporting Information for details) orientation, and the absolute charge density as a function of the distance from the graphene surface (z position in the simulation cell) in the $Na_2SO_{4,aq}$-graphene system (as illustrated in Figure 1a). Figure 1b shows that water molecules near the graphene sheets have $\theta_{DW} < 90°$ indicating that only one water OH bond points toward the graphene on average, consistent with previous studies.[5, 18, 25, 57] $\theta_{DW}$ also increases as the distance to the graphene sheet increases and stabilizes at 90° in the bulk region of the simulation cell which is the correct behavior in a homogeneous bulk environment. We hypothesize that the ability of the MLIP to correctly reproduce the lack of a net dipole in the bulk region could be due to (i) the message-passing nature of NequIP which allows interactions beyond the radial cutoff to be learned by the MLIP, (ii) interaction of the graphene sheet with water that governs organization of the interfacial water molecules, and (iii) screening of the net dipole at close to interface by subsequent water layers as also suggested recently by Cheng.[54] Consequently, the MLIP can provide a correct description of the solid-liquid interface since the orientation of water molecules near the interface is based on local interactions between the graphene sheet and the water molecules that can be captured by the MLIP. Meanwhile, Figure 1c shows the stratification of the $Na^+$ and $SO_4^{2-}$ ions with $Na^+$ ions accumulating closer (~ 5 Å) to the graphene sheets[25, 58] than the $SO_4^{2-}$ ions (~ 7 Å). Additionally, we note that in contrast to recent studies[25, 58] whereby local (short-ranged) MLIPs fail to recover charge balance in the bulk, our MLIP produces the correct charge balance in the bulk region without explicit treatment of long-range interactions despite the charge imbalanced accumulation of ions at the interface. We also note the good agreement between the MLIP-predicted and DFT-predicted bulk water density[31] (MLIP bulk water density of 0.991 g/cm$^3$ vs DFT bulk water density of 1.008 g/cm$^3$) as shown in Figure S18a, in addition to capturing the trend of the experimental bulk density with increasing $Na_2SO_4$ concentration.[59] In Figure S18b-18d, we see excellent agreement between our MLIP and DFT-predicted O-O, O-H, and H-H radial distribution function (RDF) even for distances beyond the radial cutoff of the MLIP (pink background), thus showing that the performance of the MLIP does not degrade beyond its radial cutoff. Additionally, we see good agreement between DFT and experimental RDF peak positions, further underscoring the reliability of the underlying DFT method (RPBE-D3(0)) in modeling water. As we will show later in our discussion, our MLIP reproduces experimental trends[10] in the SFG spectra, captures enhanced in-plane water diffusivity at the interface,[60] and bulk in-plane water diffusivity that is in agreement with recent computational studies[61]. Overall, the above analyses demonstrate the reliability of our MLIP in modeling the graphene-water interface in the presence of $Na_2SO_4$.

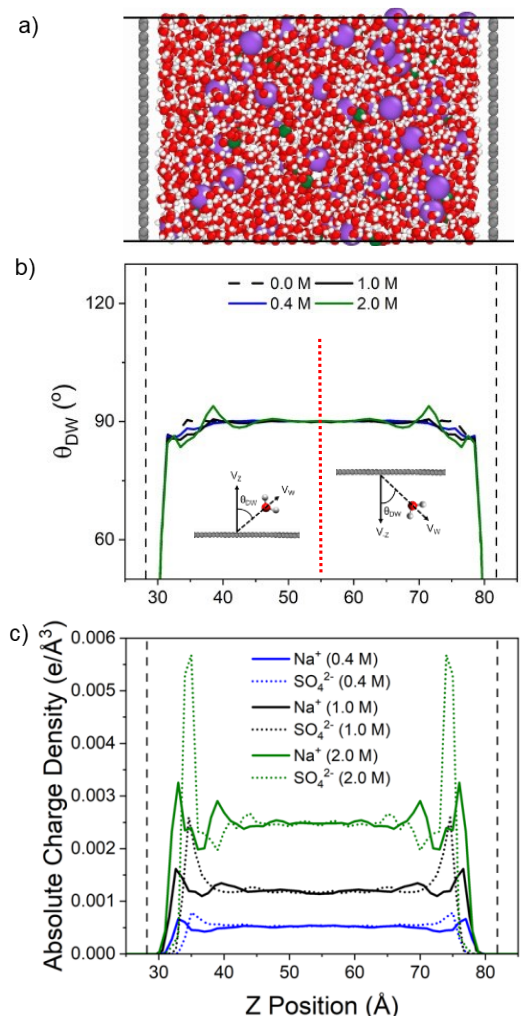

**Figure 1**. a) Snapshot of the simulated $Na_2SO_{4,aq}$ - graphene system. Distribution of b) average water dipole ($\theta_{DW}$) orientation, and c) absolute charge density as a function of z position in the simulation cell. Inset in b) shows the definition of $\theta_{DW}$ (also see Section 9 and Figure S20a of the Supporting Information). The position of the $SO_4^{2-}$ ion is taken as the position of the sulfur atom. Atom and the ion density in the z direction is taken in 1 Å bins. $Na^+$ and $SO_4^{2-}$ ions are assumed to have a +1 and -2 charge, respectively. The vertical dashed black lines indicate the position of the graphene sheets held fixed at 28.25 Å and 81.75 Å, respectively. Color code: C - gray, Na - purple, S - green, O - red, and H - white.

Next, we analyze the distribution of ions at the interface and bulk-like region in Figure 2. Figure 2a shows that $Na^+$ ions are stratified regardless of concentration as $Na^+$ ions accumulate ~5 Å from the graphene sheet while Figure 2b shows that $SO_4^{2-}$ ions prefer a more bulk-like environment ~7 Å from the graphene sheet. We rationalize $Na^+$ ions' greater proximity to graphene by noting that water molecules in the first interfacial layer are typically oriented with one OH bond pointing into the solution (see Figure 1b), and as such the oxygen end remains available to coordinate to $Na^+$ ions. Therefore, $Na^+$ ions can accumulate closer to the graphene sheet without being de-solvated (Figure S25), in agreement with previous studies of graphene-electrolyte interfaces. [7, 24, 62] As a consequence, the interface is also enriched with $SO_4^{2-}$ ions due to electrostatics without significantly changing coordination number (see Figure S25). However, we observe the enrichment of tip $SO_4^{2-}$ ion configuration at the interface (see definition and analysis in Figure S26). Previous work[10] had suggested a specific propensity of $SO_4^{2-}$ ions at the graphene-water interface, however, our simulations show that $Na^+$ ions are situated between the first (L1) and second (L2) water layers (see Section 9 of the Supporting Information for definitions), while $SO_4^{2-}$ ions are present within the L2 layer (see Figure S21b and S23b). We attribute the accumulation of $SO_4^{2-}$ ions in the L2 layer to the fact that $SO_4^{2-}$ ions have a large solvation shell[31] which prevents $SO_4^{2-}$ ions from closely approaching the graphene sheet. Figure 2c shows the variation of the surface and bulk $Na^+$ and $SO_4^{2-}$ concentrations, respectively. Here, we find that the bulk concentration scales linearly with nominal concentration for both cations and anions and our MLIP produces the correct bulk concentration of $Na^+$ relative to $SO_4^{2-}$ (see also Figure S23), accounting for charge balance. Likewise, the surface concentration of the ions also increases in a linear fashion, however the ratio of the $Na^+$ and $SO_4^{2-}$ surface concentration is always below 2 and reduces as the nominal concentration increases as seen in Figure 2d. The sub-stoichiometric and decreasing $Na^+/SO_4^{2-}$ ratio at the densely packed interface can be rationalized by the fact that $Na^+$ ions rapidly saturate the boundary between the L1 and L2 layers, while $SO_4^{2-}$ ions populate the broader L2 layer. Due to the sub-stoichiometric $Na^+/SO_4^{2-}$ ratio at the interface, the interface (~10 Å region from the graphene sheet) carries a net negative charge (nominal charge of +1 and -2 assumed for $Na^+$ and $SO_4^{2-}$ ions, respectively) which increases with increasing nominal $Na_2SO_4$ concentration as shown in Figure 2d despite $Na^+$ ions accumulating closer to the graphene and having a higher surface concentration.

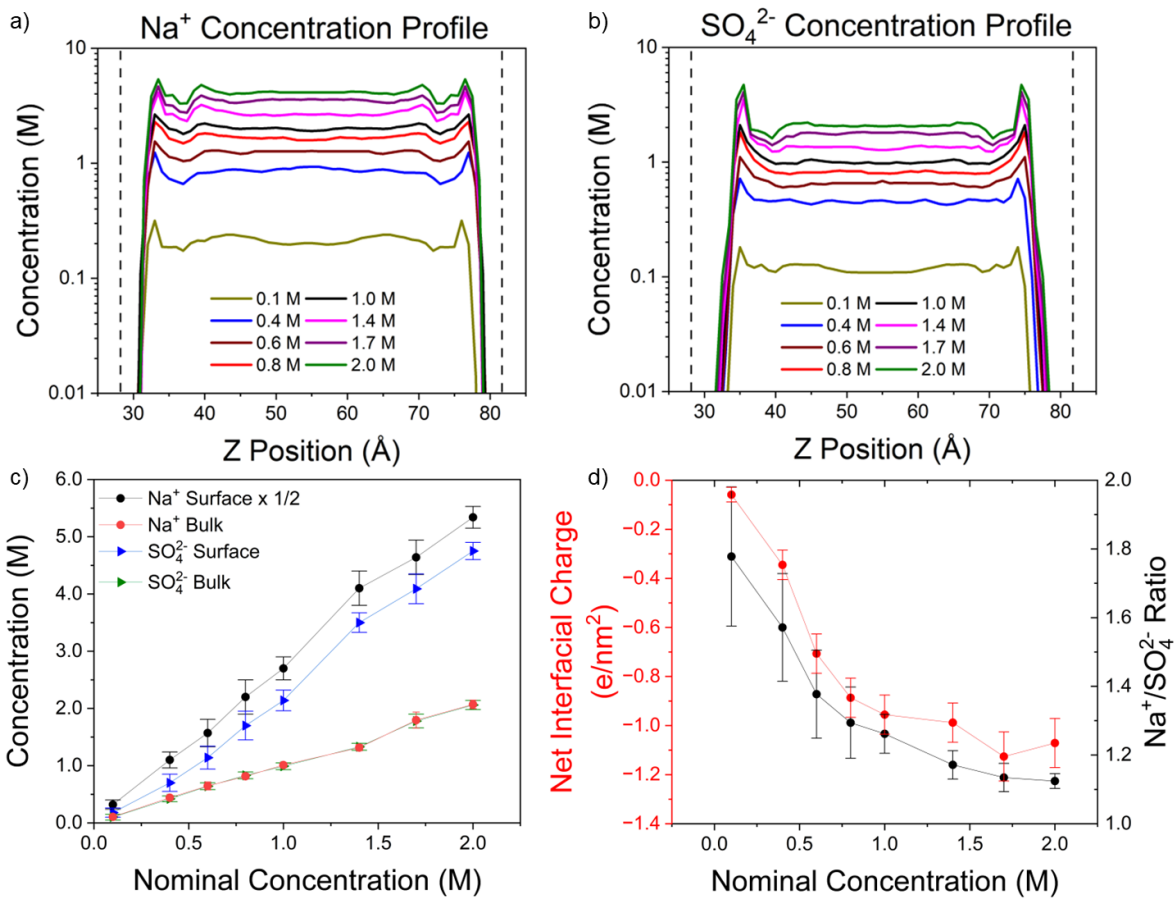

**Figure 2**. Distribution of a) $Na^+$ ions and b) $SO_4^{2-}$ ions as the function of z position in the simulation cell for various nominal concentration of $Na_2SO_4$. Note that the vertical dashed black lines represent the position of the fixed graphene sheets, and that the y-axis is on a log scale in plots a and b (See Figure S23 for plots on a linear scale and Figure S24 for a zoomed-in view of concentration profiles). The concentration profiles in a) and b) have been averaged and symmetrized over at least 50 ns of total simulation time. c) Variation of surface concentration and bulk concentration $Na^+$ and $SO_4^{2-}$ ions, respectively. The ion density in the z direction is taken in 1 Å bins. The bulk concentration of $Na^+$ ions was divided by 2 for visualization purposes (See Figure S23 for raw plot). Note that the surface concentration refers to the concentration at the position of the $Na^+$ and $SO_4^{2-}$ peaks at the interface. d) Net interfacial charge (left axis) and $Na^+/SO_4^{2-}$ ratio at the interface (right axis). The net interfacial charge is calculated as the difference between the $Na^+/SO_4^{2-}$ charge density in the combined L1, L2, and L3 water layers. Note that the $Na^+/SO_4^{2-}$ ratio is taken as the peak to peak (in concentration) ratio of the $Na^+$ and $SO_4^{2-}$ ions at the interface. Error bars represent standard deviation.

We now present our phase-resolved SFG spectra at the $Na_2SO_{4,aq}$-graphene interface across a nominal concentration range of 0.0 to 2.0 M in Figure 3. Yang et al.[10] have measured the intensity of the SFG signal (homodyne-detected SFG) for the graphene-water interface in the presence of $Na_2SO_4$ and showed an increase in the SFG peak in the hydrogen bonded region (i.e., 3000 to 3400 cm$^{-1}$) with increasing $Na_2SO_4$ concentration. However, interpreting the spectral changes is difficult because the SFG spectra lacks phase-resolution which is helpful to provide a molecular interpretation of the spectra. We therefore focus on the phase-resolved imaginary part of the simulated SFG signal and the observed trends, which can be more readily connected to a molecular and orientational interpretation, as done in recent studies.[18-20] In Figure 3a, we find that the positive peak corresponding to dangling OH bonds is relatively constant as a function of nominal $Na_2SO_4$ concentration, as also observed in experiments[10]. We attribute this to a compensating reorganization of water molecules within the L1 interfacial layer. As we show later in Figure 4, as ion concentration increases water molecules shift between two interfacial configurations that contribute comparably to the positive peak,[18] thus the intensity of the positive peak remains relatively unchanged. Additionally, we observe that the negative SFG peak in the hydrogen bonded region becomes more intense as the nominal $Na_2SO_4$ concentration increases from 0.1 M to 2.0 M, in qualitative agreement with experimental measurements,[10] confirming the reliability of our MLIP in modeling the graphene-water interface. The intensification of the negative peak with increasing $Na_2SO_4$ concentration indicates increasing orientational order at the interface that has been previously attributed to increasing interaction between water molecules and anions.[20] We also observe in Figure 3b that the negative peak position becomes red-shifted as the nominal $Na_2SO_4$ concentration increases. Since the position of the negative peak is related to the strength of the hydrogen bonds, the red-shifting of the peak in the hydrogen bonded region indicates formation of strong hydrogen bonds.[63-65] Therefore, we attribute the red-shift of the peak to the increasing number of strong hydrogen bonds between water molecules and $SO_4^{2-}$ ions[63, 64] in the interfacial region compared to the weaker hydrogen bonds between water molecules.

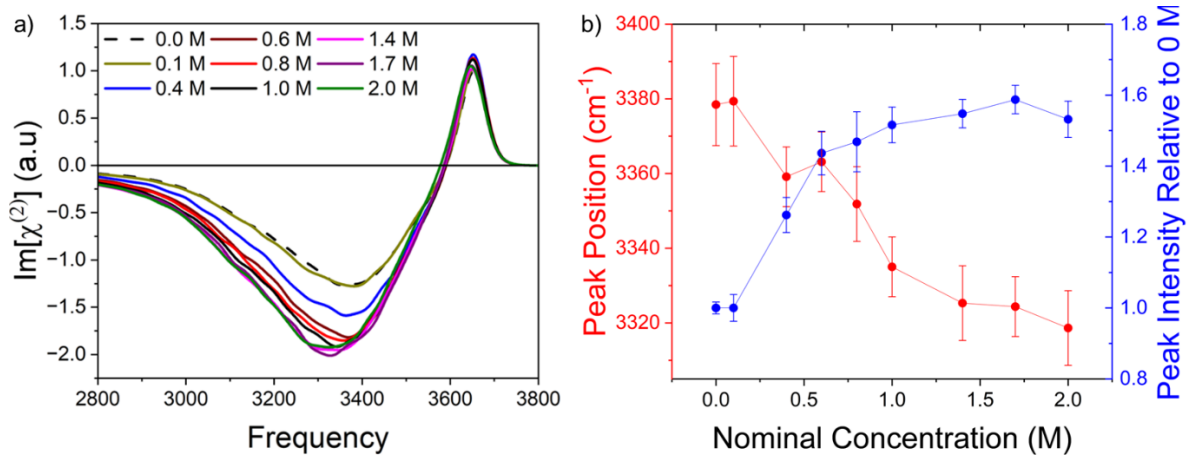

**Figure 3**. a) SFG spectra as a function of $Na_2SO_4$ nominal concentration. Note that spectra are normalized such that the dangling OH peak (at ~3650 cm$^{-1}$) at 0.0 M has an intensity of 1. Note that the frequencies have been scaled by 0.96. b) Variation of the negative peak in the hydrogen bonded region and relative intensity (compared to 0.0 M concentration) as a function of nominal $Na_2SO_4$ concentration. Note that the error bars represent standard deviation.

To rationalize the spectral changes in the hydrogen bonding region, we decompose the SFG spectra into contributions from water molecules in successive interfacial water layers as shown in Figure S27 (see Section 9 and Figure S19c of the Supporting Information for definition of interfacial water layers). We find that the first (L1) layer provides the greatest contribution to the SFG spectra, in agreement with recent computational studies;[18-20] and as concentration increases, the contribution from the L1 layer also increases, indicating enhanced organization in the L1 layer. We also observe that the contribution to the hydrogen-bonded peak from the L2 layer increases as nominal concentration increases, showing that $SO_4^{2-}$ ions accumulated in the L2 layer have a distinct impact on the water orientation. We highlight that there is no contribution to the SFG spectra from the bulk region across the studied concentrations and no meaningful contribution from the L3 layer for concentrations below 1.4 M. Above 1.4 M concentration, we observe some contribution which we attribute to enrichment of $Na^+$ ions in the L3 layer (see Figure 1c) which tends to disrupt the water structure (Figure 1b).

To further explain the effect of the ions on the SFG spectra, we now analyze the orientation of water molecules within the interfacial water layers. Previous studies have suggested that an increase in thickness of the interfacial water layers leads to an intensification of the SFG spectra in the low frequency hydrogen bonded region (3000 to 3400 $cm^{-1}$).[66-68] However, the thickness of the interfacial water layers (L1, L2, L3) remain relatively unchanged over the range of the studied concentrations as we show in Figure S19. Meanwhile, other studies have attributed the enhancement of the peak in the hydrogen bonding region to an increase in water molecules having a "downward" orientation into the bulk of the solution.[18-20, 69] Therefore, we analyze the orientation of water molecules in the interfacial water (L1, L2, and L3) layers via the joint probability density distribution $P(\cos\theta_{Dw}, \cos\theta_{HH})$ as described in Figure 4 and the Supporting Information (see Section 9 and Figure S20 of the Supporting Information).

Figure 4 below shows the difference plots between the L1 layer $P(\cos\theta_{Dw}, \cos\theta_{HH})$ at the 0.1, 0.4, 1.0, and 2.0 M concentrations and the L1 layer $P(\cos\theta_{Dw}, \cos\theta_{HH})$ at 0.0 M concentration (i.e., neat water). At 0.1 M concentration (Figure 4a), there is only a minimal increase in density in the VA configuration (Figure 4e) and no change elsewhere. Meanwhile as concentration increases (Figure 4b to 4d), we observe that the L1 layer becomes more structured as shown in the increase in density in the VA configuration. It should be noted that previous studies[18] have also shown that the VA contribution contributes to both the positive and negative features of the SFG spectra and thus the spectral changes in the hydrogen bonded region, with increasing $Na_2SO_4$ concentration, are consistent with our orientational analysis and layer-by-layer SFG analysis. Likewise, we find that the L2 layer becomes more structured as $Na_2SO_4$ concentration increases (0.1 to 2.0 M) as shown by increase in the density of the BA configuration (Figure S31) which is known to enhance the SFG spectra in the hydrogen bonded region.[18, 19, 69] Therefore, we attribute the increasing contribution from the L2 layer to the increasing population of water molecules adopting the BA configuration induced by $SO_4^{2-}$ ions in the L2 layer. Meanwhile in the L3 layer, changes in the orientation of water molecules are non-monotonic below 1.4 M concentration. However, above 1.4 M concentration the BT orientation is enriched as $Na^+$ ions also enrich the L3 layer, thereby leading to positive contributions to the SFG spectra in the hydrogen bonded region.[18]

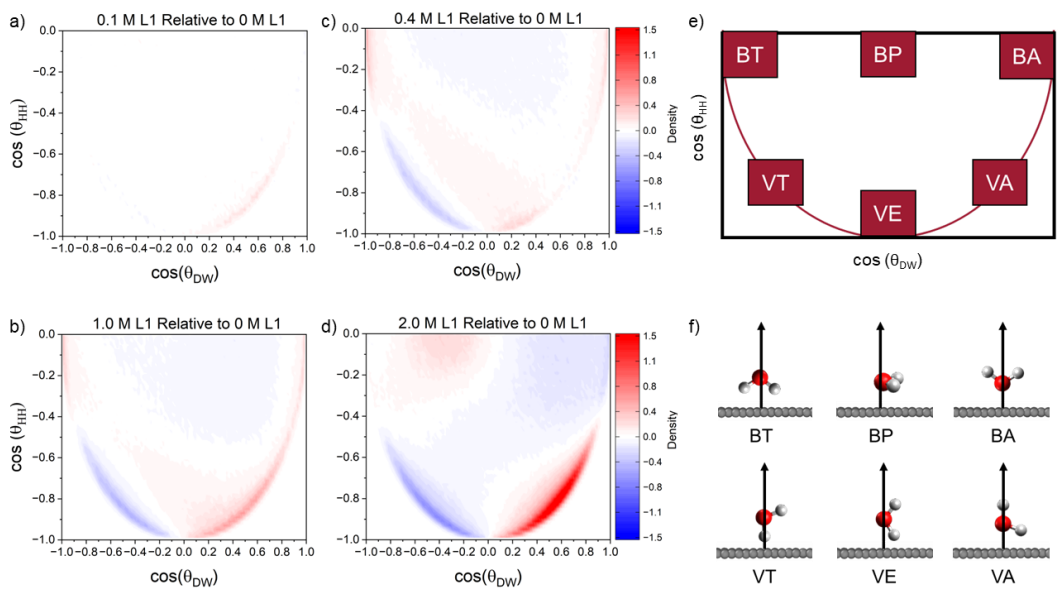

**Figure 4**. Joint probability density distribution of the water orientation ($P(\cos\theta_{Dw}, \cos\theta_{HH})$) in the L1 layer at a) 0.1 M, b) 0.4 M, c) 1.0 M, and d) 2.0 M nominal $Na_2SO_4$ concentration. e & f) Definition of water orientations (see Section 9 of Supporting Information for further details). See Figure S28 for raw plots, and Figure S29 for the difference plots for all other concentrations.

To further elucidate the structure of water within the interfacial layers, we also analyzed the hydrogen bonding by computing the number of hydrogen bonds/water molecules in the water layers as a function of nominal $Na_2SO_4$ concentration (Figure 5a). Here, we observe that the ions do not influence the hydrogen bonding at 0.1 M concentration, which is similar to our observations that the water structure in the L1 layer is relatively unperturbed by ions at 0.1 M concentration. At 0.4 M concentration, we find that the number of hydrogen bonds/water increases and subsequently reduces as concentration increases. In the L2 layer, L3 layer, and bulk we also observe that in general the number of hydrogen bonds/water is highest at 0.0 M concentration and reduces as concentration increases and that the number of hydrogen bonds/water is close to the bulk value outside the L1 layer. Additional insights into the structure of water in the interfacial layers can be gained by analyzing the underlying hydrogen bonding network. Here, we analyze contributions from single donor-single acceptor (1D-1A), double donor-single acceptor (2D-1A), single donor-double acceptor (1D-2A), and double donor-double acceptor (2D-2A) water molecules (see Section 13 of the Supporting Information). In the L1 layer, the HB distribution at 0.1 M closely resembles 0.0 M, dominated by the 1D-1A (~28.5%) and 2D-1A (~21.7%) configurations as shown in Table S2. Meanwhile, at 0.4 M concentration there are more fully hydrogen bonded water molecules (i.e., 2D-2A = 23.4%) which lead to an increase in the number of hydrogen bonds/water as shown in Figure 5a. With further concentration increase, ordered configurations (1D-1A, 1D-2A, 2D-1A, 2D-2A) progressively give way to distorted ones (1D-0A, 0D-1A, 0D-0A, 0D-2A, 2D-0A), reducing the overall hydrogen bonds per water molecule (see Section 13 of Supporting Information for additional discussion). Therefore, the increase in distorted hydrogen bonding configurations with increasing $Na_2SO_4$ concentration is reflective of the increasing ion-induced re-orientation of water molecules at the interface (see Section 13 of the Supporting Information for additional details).

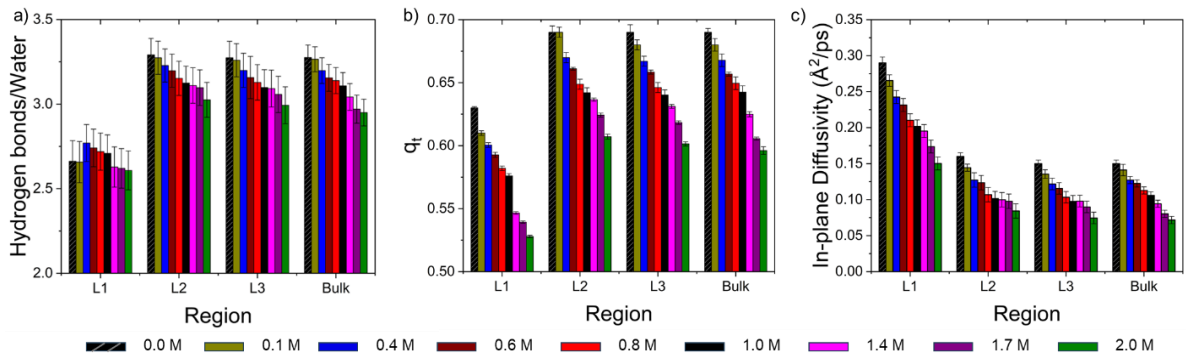

**Figure 5**. a) Hydrogen bonds per water, b) Tetrahedral order parameter ($q_t$), and c) in-plane (2D) water diffusivity as a function of nominal $Na_2SO_4$ concentration in the L1, L2, L3, and bulk water layers. Note that the error bars represent standard deviation.

We also explored the tetrahedral order parameter ($q_t$) and the diffusivity of water molecules in the interfacial and bulk water layers (Figure 5b and 5c). We see in Figure 5b that $q_t$ decreases monotonically from 0.0 to 1.0 M in L1; however, above 1.0 M there is a sharp drop in $q_t$ which we attribute to the sharp increase in the surface concentrations of $Na^+$ and $SO_4^{2-}$ ions above 1.0 M nominal concentration as shown in Figure 2a and 2b. In general, $q_t$ reduces with increasing $Na_2SO_4$ concentration in the L2 and L3 layers and bulk. Figure 5c shows that water molecules in the L1 layer have the highest 2D diffusivity which is a well-known consequence of the reduced hydrogen bonding at the interface,[56, 70, 71] while our predicted bulk in-plane diffusivity at 0.0 M concentration is also consistent with recent studies.[61] It is also generally true that the increasing concentration of ions reduces the water diffusivity in each layer. Lastly, we estimate the water 3D diffusivity in bulk as a function of $Na_2SO_4$ concentration and system size (i.e., number of water molecules) as shown in Figure S35. In agreement with our previous work,[31] we find that at each corresponding system size, the water diffusivity decreases gradually. Our analysis shows that increasing concentration of ions breaks the tetrahedral order of water, while increasing interaction with the ions tends to reduce the water diffusivity as $Na_2SO_4$ concentration increases.

**Conclusions**

In this computational study, we used MLIP-based molecular simulations and simulated SFG spectroscopy to study the structure of the $Na_2SO_{4,aq}$-graphene interface and provide a molecular interpretation of previous experimental SFG spectra. We also resolved the atomistic structure of the $Na_2SO_{4,aq}$-graphene interface and provide the molecular origins of the apparent discrepancy between the recent suggestion[10] of $SO_4^{2-}$ enrichment at the $Na_2SO_{4,aq}$-graphene interface. We showed that preferential $SO_4^{2-}$ enrichment at the $Na_2SO_{4,aq}$-graphene interface and cation accumulation at the interface are not mutually exclusive. Our results showed that the ions are stratified, with $Na^+$ ions accumulating between the first and second interfacial water layers (~ 5 Å from the graphene sheet), while $SO_4^{2-}$ ions accumulate within the second interfacial water layer (~ 7 Å from the graphene sheet), thus emphasizing that Na cations accumulate near graphene-water interfaces over sulfate anions. Our analysis indicated that the interfacial region (10 Å from the graphene sheet) has a net negative charge due to the sub-stoichiometric $Na^+/SO_4^{2-}$ ratio at the interface, rationalizing the previous observation of a negatively charged interface.[10] In qualitative agreement with previous experimental work, we showed that the SFG peak in the hydrogen bonding region intensifies as a function of $Na_2SO_4$ concentration. By further analyzing the layer-by-layer SFG contributions and water orientation, we demonstrated that the enhancement of the SFG peak in the hydrogen bonded region is due to increasing contribution from the $SO_4^{2-}$-rich second interfacial water layer in which $SO_4^{2-}$ ions orient water molecules such that their OH bonds point into the solution. Our work highlights the necessity of integrating SFG measurements with molecular simulations to obtain an atomistic resolution of the interfacial water structure and emphasizes the crucial role of MLIPs that allow us to maintain DFT-level accuracy while achieving ns-long simulations that are far beyond the current accessible timescales of AIMD. By resolving the composition of the interface and ion-induced reorganization, MLIPs provide a means of gaining deeper insights beyond the reach of continuum models into interfacial phenomena that will be important in resolving open questions related to electrolyte-graphene interfaces.

Overall, we demonstrated the applicability of MLIPs in capturing trends in the experimental SFG spectra, due to increasing $Na_2SO_4$ concentration, and provided molecular-level interpretation. Our MLIP-based approach provides further evidence for the reliability of MLIPs in modeling interfacial systems and opens the possibility of further development of MLIPs to model complex solid-liquid interfaces and interfacial phenomena such as reactions. Insights into the structure of the electrolyte-graphene interface are not only critical for a better understanding of interfacial phenomena in the presence of ions but also provide avenues for the design of more performant and stable electrochemical systems under various conditions.

## Supporting Information

Figures S1–S35, which present parity and error distribution analyses for MLIP-predicted energies, and forces; correlation between committee uncertainty and true error; distributions of interatomic distances, bond lengths, and coordination numbers in both training and test sets; definition of water layers and orientations; net charge density as a function of z position; ion distribution as a function of z position, and $Na^+/SO_4^{2-}$ bulk and surface concentrations as a function of nominal $Na_2SO_4$ concentration; ion solvation structure as a function of z position; $SO_4^{2-}$ orientation relative to graphene sheet as a function of z position; layer-by-layer decomposition of the SFG spectra; layer-by-layer $P(\cos\theta_{Dw}, \cos\theta_{HH})$ distributions; 3D water diffusivity in bulk; Tables S1–S5 summarizing sodium and sulfate ion concentration at the surface and in the bulk of the solution; layer-by-layer distribution of hydrogen bonding configurations as a function of $Na_2SO_4$ concentration (PDF)

Full training, validation, and test sets and input files for training the NequIP models can be downloaded from our Zenodo repository (DOI: https://doi.org/10.5281/zenodo.15512423).

## Acknowledgements

This work was also made possible by the U.S. Department of Energy, Office of Science, CPIMS program, under Award DE-SC0024654. The authors thank Tristan Maxson, Sophia Ezendu, Gbolagade Olajide, and Mustapha Iddrisu for their insightful comments on the manuscript and work. This work was also made possible in part by a grant of high-performance computing resources and technical support from the Alabama Supercomputer Authority. This research used computational resources of the National Energy Research Scientific Computing Center (NERSC), a U.S. Department of Energy Office of Science User Facility located at Lawrence Berkeley National Laboratory, operated under Contract DE-AC02-05CH11231 using NERSC Award BES-ERCAP0024218. Any opinions, findings, conclusions, and/or recommendations expressed in this material are those of the authors(s) and do not necessarily reflect the views of funding agencies.